\makeatletter

\def\CLASSINPUTinnersidemargin{0.2in}%
\def\CLASSINPUToutersidemargin{0.2in}%
\def\CLASSINPUTtoptextmargin{0.2in}%
\def\CLASSINPUTbottomtextmargin{0.2in}%

\makeatother
\documentclass[conference]{IEEEtran}
\IEEEoverridecommandlockouts
\usepackage{cite}
\usepackage{amsmath,amssymb,amsfonts}
\usepackage{algorithmic}
\usepackage{graphicx}
\usepackage{textcomp}
\usepackage{xcolor}
\usepackage{cite}
\usepackage{amsmath,amssymb,amsfonts}
\usepackage{algorithmic}
\usepackage{graphicx}
\usepackage{textcomp}
\usepackage{xcolor}
\usepackage{multirow}
\usepackage{scrextend}
\usepackage{colortbl}
\usepackage{hhline}
\usepackage{footnote}
\makesavenoteenv{tabular}
\usepackage{booktabs}
\usepackage{adjustbox}
\usepackage{tabularray}
\usepackage{algorithm}
\usepackage{float}
\usepackage{tikz,graphics,color,fullpage,float,epsf,caption,subcaption}
\usepackage{caption}
\usepackage{subcaption}
\usepackage{xspace}
\usepackage{enumitem}
\usepackage{booktabs}
\usepackage{multirow}
\usepackage{fancyhdr}
\newcommand*{\affaddr}[1]{#1} 
\newcommand*{\affmark}[1][*]{\textsuperscript{#1}}

\def\BibTeX{{\rm B\kern-.05em{\sc i\kern-.025em b}\kern-.08em
    T\kern-.1667em\lower.7ex\hbox{E}\kern-.125emX}}
\begin{document}
\setlength{\belowdisplayskip}{4pt} \setlength{\belowdisplayshortskip}{4pt}
\setlength{\abovedisplayskip}{4pt} \setlength{\abovedisplayshortskip}{4pt}
\title{Video Quality Assessment for Resolution Cross-Over in Live Sports}

\author{Jingwen Zhu\affmark[1,2], Yixu Chen\affmark[2], Hai Wei\affmark[2], Sriram Sethuraman\affmark[2], Yongjun Wu\affmark[2]
\\\affaddr{\affmark[1]Nantes Université, Ecole Centrale Nantes, CNRS, LS2N, UMR 6004, Nantes, France} \\ \affaddr{\affmark[2]Amazon Prime Video, Seattle, USA}  } 
\maketitle

\begin{abstract}
In adaptive bitrate streaming, resolution cross-over refers to the point on the convex hull where the encoding resolution should switch to achieve better quality.
Accurate cross-over prediction is crucial for streaming providers to optimize resolution at given bandwidths.
Most existing works rely on objective Video Quality Metrics (VQM), particularly VMAF, to determine the resolution cross-over. However, these metrics have limitations in accurately predicting resolution cross-overs. Furthermore, widely used VQMs are often trained on subjective datasets collected using the Absolute Category Rating (ACR) methodologies, which we demonstrate introduces significant uncertainty and errors in resolution cross-over predictions.
To address these problems, we first investigate different subjective methodologies and demonstrate that Pairwise Comparison (PC) achieves better cross-over accuracy than ACR. We then propose a novel metric, Resolution Cross-over Quality Loss (RCQL), to measure the quality loss caused by resolution cross-over errors. 
Furthermore, we collected a new subjective dataset (LSCO) focusing on live streaming scenarios and evaluated widely used VQMs, by benchmarking their resolution cross-over accuracy. 

\end{abstract}


\section{Introduction}
Video streaming has experienced significant growth, driven by the widespread availability of high-speed Internet and the increasing use of mobile devices. It has become an essential part of daily life, serving purposes in entertainment, education, business, and more.
To accommodate varying bandwidth and device types among end-users, Adaptive BitRate streaming (ABR) methods~\cite{bentaleb_survey_2019} are widely used. In ABR, video content is encoded at multiple bitrate-resolution pairs, referred to as representations. These representations form a bitrate ladder~\cite{amirpour_pstr_2021}, allowing video quality to be dynamically adjusted based on the viewer's available bandwidth and device capabilities.

\begin{figure}[t]
    \centering
    \includegraphics[clip,width=0.35\textwidth]{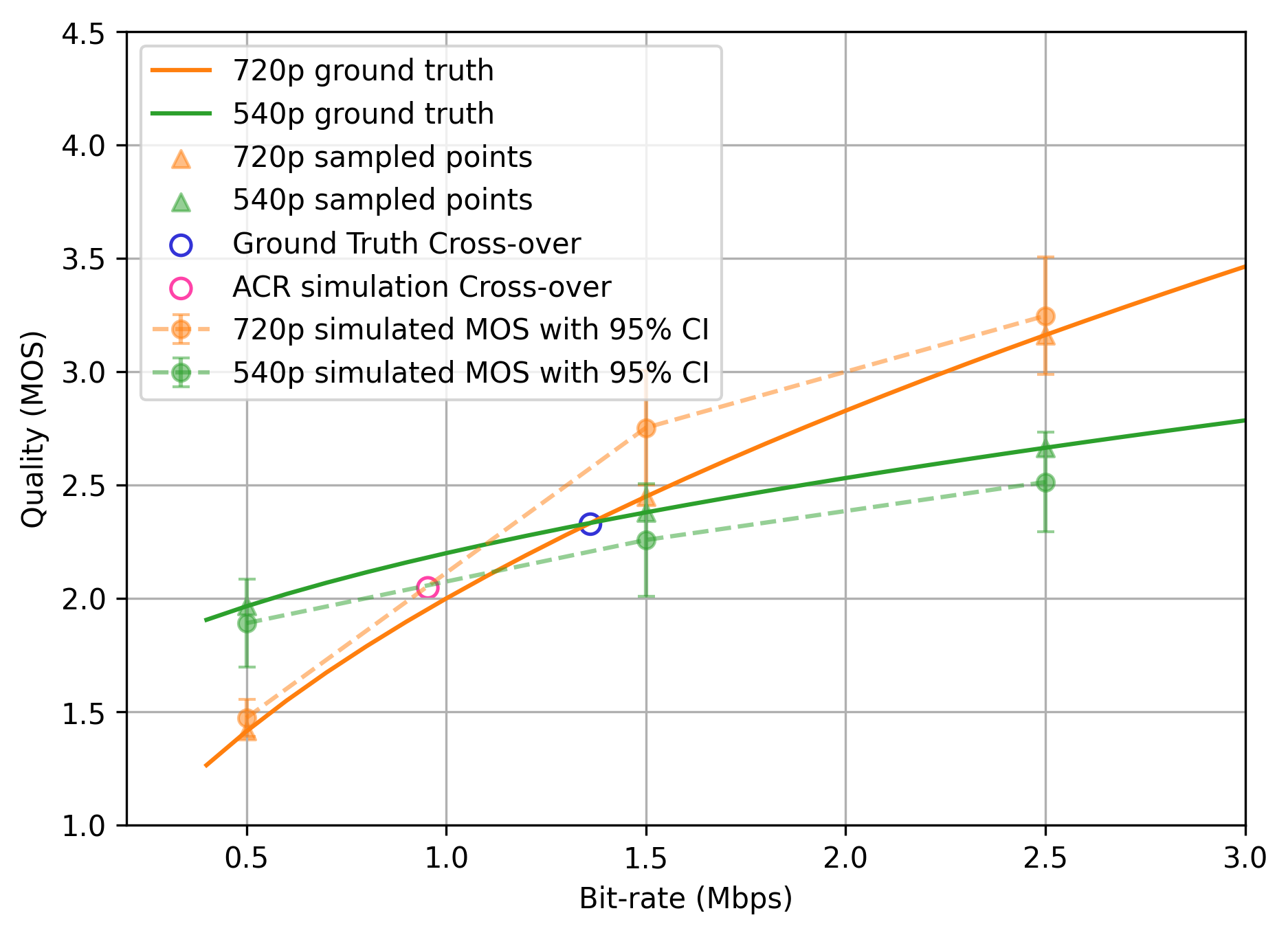}
    \caption{Illustration of resolution cross-over error introduced by ACR: Solid lines show the ground truth cross-over at ~1.4 Mbps between 720p and 540p. Dashed lines represent MOS from ACR simulations, with the cross-over at ~900 kbps.}
    \label{fig:teaser}
    \vspace{-0.5cm}
\end{figure}

Fixed bitrate ladders, often used in streaming protocols like HLS, have traditionally been employed.
However, this ``one-size-fits-all" approach may not be optimal due to the diversity of video content, and it often fails to deliver the best possible quality to end users. To address this, many studies have proposed per-title or per-scene bitrate laddering~\cite{menon2023just, katsavounidis2021iterative,durbha2024bitrate,  srikar2024constructing}. In these approaches, videos are encoded with varying parameters, such as resolution and bitrate, and their quality is evaluated. An optimized bitrate ladder is then constructed by selecting representations from a convex hull derived from the quality measurements of the encoded representations.

Resolution cross-over refers to the point on the convex hull where the encoding resolution should switch to achieve better quality. The solid orange and green lines in Fig.~\ref{fig:teaser} illustrate the resolution cross-over between 720p and 540p. 
For this video content, if the bitrate is higher than the cross-over point (1.4 Mbps), the representations with 720p encoding resolution will be preferred over 540p. In contrast, 540p representation should be used.

Previous works have attempted to predict the resolution cross-over directly~\cite{bhat2021combining} or to use the resolution cross-over to design the bitrate ladder~\cite{menon2023just, katsavounidis2021iterative,durbha2024bitrate,  srikar2024constructing}. These studies primarily rely on Video Quality Metrics (VQMs). While it is well understood that subjective quality assessments are more reliable, they are often too expensive and impractical to perform for every piece of video content, particularly in live streaming scenarios. Consequently, these works use objective VQMs, especially the perceptual quality metric VMAF~\cite{vmaf}, as a pseudo ``ground truth" for determining the resolution cross-over.

However, while VQMs such as VMAF and P1204.3~\cite{rao2020bitstream} have shown strong correlations with subjective video quality, their accuracy in predicting the resolution cross-over compared to subjective assessments remains an open question. Chen \textit{et al.}~\cite{chen2024encoder} demonstrated that both VMAF and P1204.3 often fail to predict the resolution cross-over accurately and tend to incorrectly favor higher-resolution videos.

Furthermore, learning-based objective VQMs, such as VMAF, P1204.3, EQM~\cite{chen2024encoder} are trained and tested on subjective datasets collected using the Absolute Category Rating (ACR) methodology~\cite{vmaf, rao2020bitstream, chen2024encoder, hdr_sdr, lbs, hdr_sports}. As shown in Fig.~\ref{fig:teaser}, our simulation demonstrates that due to the inherent uncertainty in ACR subjective studies, the resolution cross-over derived from ACR studies can significantly deviate from the ground truth assumption (detailed in~\ref{sec:acr}).

Moreover, to the best of our knowledge, no established metric exists to evaluate the accuracy of resolution cross-over obtained by VQM. For VQM evaluation, the most commonly used metrics are correlations, such as Spearman Rank Order Correlation Coefficient (SROCC) and Pearson Linear Correlation Coefficient (PLCC). However, it is evident that a higher correlation with subjective datasets does not necessarily guarantee improved resolution cross-over accuracy.

This paper addresses the challenges associated with resolution cross-over prediction in the context of live streaming by presenting the following contributions:
\begin{enumerate}
\item We demonstrate errors in resolution cross-over determination using ACR through simulation.
\item A pilot study comparing ACR and PC shows that PC provides more accurate resolution cross-over.
\item We introduce Resolution Cross-Over Quality Loss (RCQL) to assess cross-over accuracy.
\item We collect the Live Sport Cross-Over (LSCO) dataset, propose a method for cleaning incomplete PC data, and benchmark RCQL on LSCO for live streaming use cases.
\end{enumerate}

\section{Subjective study design for resolution cross-over}
\subsection{ACR and its limitation}
\label{sec:acr}
ACR is a widely used methodology for video quality evaluation. Participants rate video stimuli on a predefined scale, often with a hidden reference (ACR-HR). A common example is the 5-point scale: 5 (Excellent) to 1 (Bad), with variants like 5-point or 9-point discrete/continuous scales~\cite{huynh2010study}. ACR is popular for being easy to conduct and time-efficient, with results comparable to other methods like DSIS and SAMVIQ~\cite{tominaga2010performance}. However, ACR has limitations, \textit{e.g.,} subjects may interpret the scale differently, leading to inconsistencies.

We assessed the precision of ACR in determining resolution cross-over points through simulations based on real data. Ground truth quality values were manually chosen as plausible assumptions. Observer ratings $R_s$ deviate from these assumptions (ground truth $\mu_s$), following a Gaussian distribution with standard deviation $\sigma_s$, influenced by quality range and test environment, as per the SOS hypothesis~\cite{hossfeld2011sos}.
Using the HDR-LIVE dataset~\cite{shang2023study}, we modeled $\sigma_s$ via the SOS curve.
Adding $\sigma_s$ to the ground truth assumptions allowed us to simulate ACR ratings for $N$ participants ($N$=33 to be consistent with HDR-LIVE) by sampling from this distribution. The simulated Mean Opinion Score (MOS) was then obtained by averaging these simulated ratings. Fig.~\ref{fig:teaser} shows that due to ACR uncertainty, the cross-over points can significantly deviate from the ground truth.

\subsection{Pair Comparison with Active Sampling}
Considering the errors introduced by ACR for determining the cross-over point, we decided to use alternative test methodologies. Perez-Ortiz \textit{et al.}~\cite{perez2017practical} demonstrated that PC is more accurate method compared to ACR-HR. Similarly, Testolina \textit{et al.}~\cite{testolina2024fine} proposed using PC for fine-grained subjective visual quality assessment. We therefore considered PC to collect subjective data for resolution cross-over evaluation.


One major limitation of PC is its $O(n^2)$ time complexity~\cite{li2018hybrid}, where $n$ is the number of candidate stimuli. Several methods~\cite{li2013boosting, ponomarenko2015image,li2018hybrid, mikhailiuk2021active} have been proposed to reduce the number of comparisons, among which Active Sampling achieves the best performance~\cite{mohammadi2022evaluation}.
We used the Active Sampling method from~\cite{mikhailiuk2021active} for dataset collection. Active Sampling dynamically selects the most informative pairs for comparison instead of evaluating all combinations. The PC results are stored as a Pair Comparison Matrix (PCM). To convert the PCM into a continuous quality scale, we used the pwcmp tool~\cite{perez2017practical} to recover the Just Objectionable Difference (JOD) scale~\cite{perez2019pairwise}. This approach relies on Thurstone Case \textit{V} assumptions, calibrated so a 1-unit difference on the quality scale corresponds to 75\% of observers selecting one video over the other. The JOD reconstruction is performed per video content, with no cross-content comparisons.

\begin{figure}[t]
    \centering
    \begin{subfigure}{.24\textwidth}
        \centering
        \includegraphics[width=\textwidth]{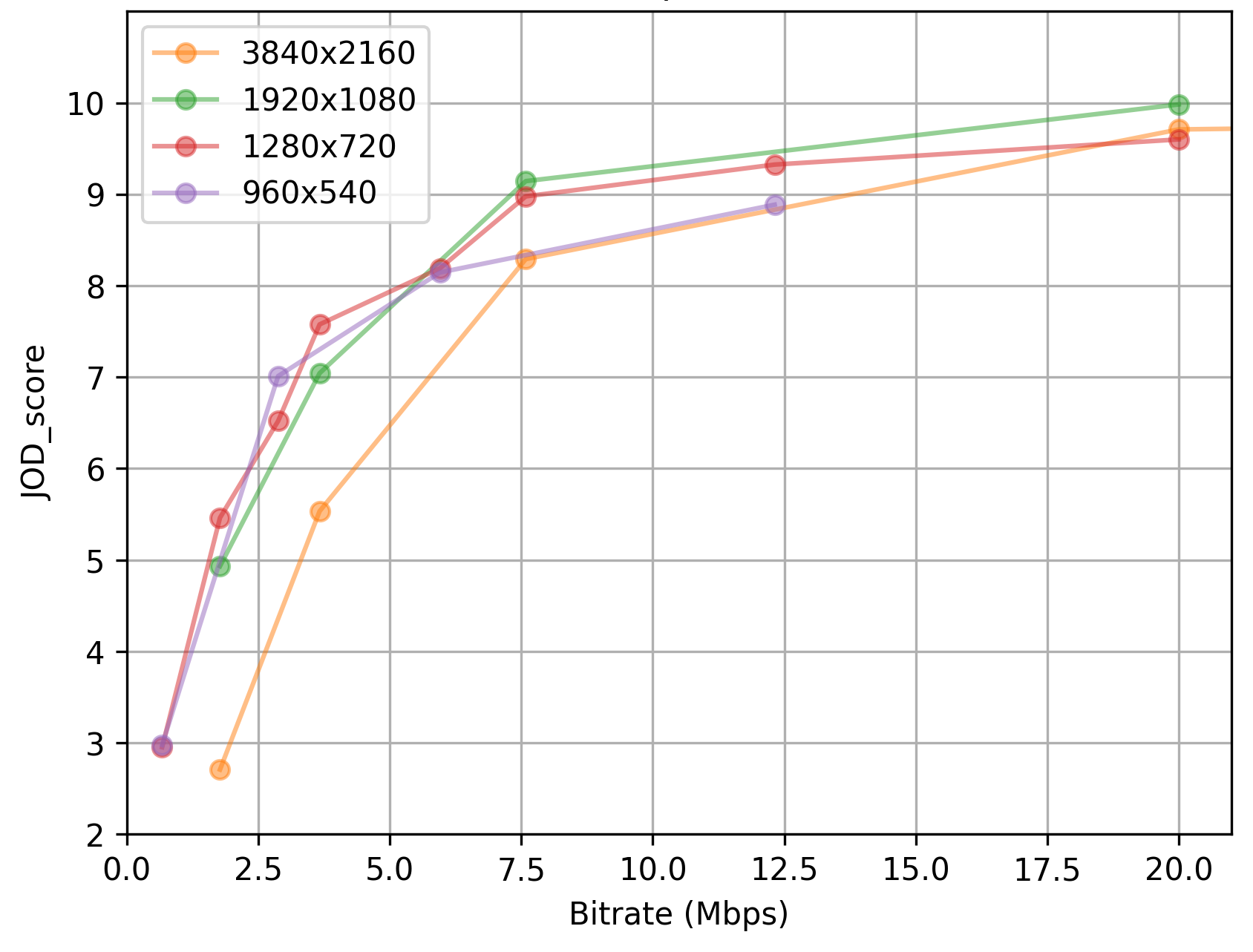}
        \vspace{-0.5cm}
        \caption{PC of EPL7}
        \label{subfig:epl7_pc}
    \end{subfigure}
    \begin{subfigure}{.24\textwidth}
        \centering
        \includegraphics[width=\textwidth]{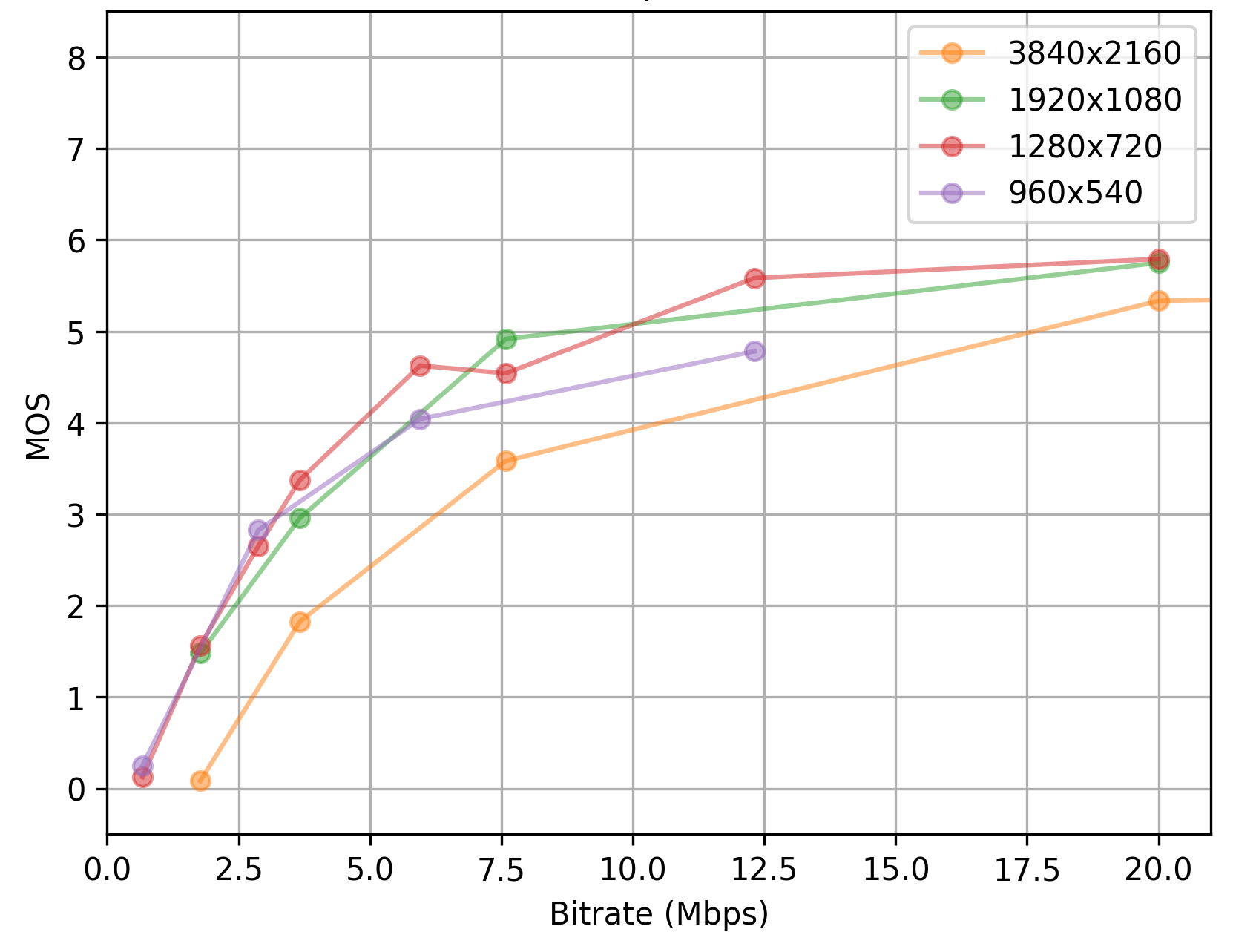}
        \vspace{-0.5cm}
        \caption{ACR of EPL7}
        \label{subfig:epl7_acr}
    \end{subfigure}
    \caption{Pilot study example: Comparison of PC vs ACR}
    \vspace{-0.5cm}
    \label{fig:pilot}
\end{figure}

To further confirm that PC can provide more accurate cross-over than ACR, we conducted a pilot study using both ACR, PC with active sampling and an expert viewing evaluation. For ACR, we used a 9-point discrete scale~\cite{itu2022BT910}. A MOS value of 0 corresponds to ``Bad", and a MOS value of 8 corresponds to``Excellent". Three 10-second video clips from live streaming applications were selected and encoded at resolutions of 2160p, 1080p, 720p, and 540p, with bitrates ranging from 200 Kbps to 20 Mbps, resulting in 67 encoded videos. The study involved 24 participants for ACR, resulting in a total of 24x67 = 1,608 ratings. For PC, the number of active sampling iterations was set to 25~\cite{mikhailiuk2021active}, producing 1,650 ratings (30 participants × 55 ratings each), which is comparable to the number of ratings in ACR.
Fig.~\ref{fig:pilot} showcases the MOS from ACR and the JOD from PC obtained in the pilot study for video content EPL7. (Continuous curves are generated using `pchip' interpolation.) It can be observed that the cross-over points derived from PC and ACR differ significantly.
Five experts participated in the expert viewing, where they were asked to indicate their preferred resolution for each bitrate of each content in the dataset. The results are presented in Table~\ref{tab:expert}. An ``x" indicates that no cross-over was observed within the specified bitrate range (200 Kbps to 20 Mbps). For cases with multiple cross-over points, such as the cross-over between 1080p and 720p in EPL7 with ACR (Fig.~\ref{subfig:epl7_acr}), we computed the average by taking the mean of the minimum and maximum cross-over bitrates.
The cross-over bitrates obtained using PC align more closely with expert evaluations than those from ACR. PC data also shows better correlation with bitrate within the same resolution, following the principle that lower bitrates should not result in better quality within same resolution for the same encoding settings. Table~\ref{tab:monotonicity} presents SROCC results, indicating PC achieves better correlation in most cases.
\begin{table}[htb]
\caption{PC and ACR Pilot Study Results Compared with Expert Viewing. The values are in Mbps.}
\vspace{-0.2cm}
\label{tab:expert}
\begin{adjustbox}{width=0.8\columnwidth,center}
\begin{tabular}{@{}c|cccc@{}}
\toprule
                          &        & 4k vs 1080p    & 1080p vs 720p & 720p vs 540p  \\ \midrule
\multirow{3}{*}{EPL7}     & Expert & x              & 5.95          & 4.67          \\
                          & PC     & x              & \textbf{5.66} & \textbf{2.85} \\
                          & ACR    & x              & 8.47          & 2.23          \\ \midrule
\multirow{3}{*}{UCL8}     & Expert & 11.87          & 4.29          & 1.39          \\
                          & PC     & \textbf{12.67} & \textbf{3.80} & 0.84          \\
                          & ACR    & 19.46          & 2.37          & \textbf{1.92} \\ \midrule
\multirow{3}{*}{tennis 5} & Expert & 7.19           & 1.09          & 1.39          \\
                          & PC     & \textbf{6.59}  & \textbf{1.47} & x             \\
                          & ACR    & 14.00          & 1.77          & \textbf{0.82} \\ \bottomrule
\end{tabular}
\end{adjustbox}
\end{table}

\begin{table}[htb]
\caption{SROCC between bitrate and quality scores for ACR and PC in the pilot study.}
\vspace{-0.2cm}
\label{tab:monotonicity}
\begin{adjustbox}{width=0.7\columnwidth,center}
\begin{tabular}{@{}l|lllll@{}}
\toprule
                          &     & 2160p           & 1080p           & 720p            & 540p            \\ \midrule
\multirow{2}{*}{EPL7}     & PC  & 1.0000          & 1.0000          & \textbf{1.0000} & 1.0000          \\
                          & ACR & 1.0000          & 1.0000          & 0.9762          & 1.0000          \\ \midrule
\multirow{2}{*}{UCL8}     & PC  & 0.9000          & \textbf{1.0000} & \textbf{0.9762} & 1.0000          \\
                          & ACR & 0.9000          & 0.9429          & 0.9701          & 1.0000          \\ \midrule
\multirow{2}{*}{tennis 5} & PC  & \textbf{1.0000} & 0.9643          & \textbf{1.0000} & 0.8000          \\
                          & ACR & 0.9000          & 0.9643          & 0.9910          & \textbf{1.0000} \\ \bottomrule
\end{tabular}
\end{adjustbox}
\vspace{-0.5cm}
\end{table}

\subsection{LSCO Dataset collection}

Based on the results from the pilot study, we conducted a full study for resolution cross-over evaluation using PC with active sampling. The collected dataset is named Live Sport Cross-Over (LSCO).We selected 20 video clips covering a wide range of Spatial Information (SI) and Temporal Information (TI)~\cite{itu2022BT910}, each approximately 10 seconds long, focusing on live sports events such as football, soccer, and tennis.
The selected content was encoded using our in-house encoder into different resolutions, similar to the setup in the pilot study.
To minimize participant fatigue, each participant evaluated only 55 pairs, and each session lasted less than 30 minutes. A total of 131 participants contributed to the study, resulting in over 7,000 collected ratings.
We used two calibrated OLED55C3PUA 4K displays, with the viewing distance controlled at 1.5 times the screen height following the ITU recommendation~\cite{itu2022BT910}. 

\section{Observer screening}
\label{sec:obs_screen}
Observer screening is crucial for reliable QoE datasets. While well-established methods exist for ACR and DCR~\cite{itu2022BT910, zhu2023zrec}, screening for PC data is less developed. Methods like Cohen’s Kappa~\cite{cohen1960coefficient} and RT distance~\cite{rogers1960computer}, and Ak \textit{et al.}'s ambiguity-weighted RT distance~\cite{ak2022rv}, are limited to complete PC data.
We propose a new approach to compute inter-observer consistency for incomplete PCMs collected via active sampling. This metric helps identify and filter outliers by considering: 1) pair ambiguity, 2) observer agreement, and 3) the number of ratings per pair.
\subsection{Ambiguity}
For each observer $o_i$, assume that he/she voted on $N$ pairs of videos, with each pair denoted as $p_n$. The number of ratings (i.e., how many people voted on this pair) for $p_n$ is represented as $r_n$.
For pair $p_n$, the number of ratings for $A$ is denoted as $a_n$, the number of ratings for $B$ is denoted as $b_n$, and the number of ties is denoted as $t_n$. We know that:
\begin{equation}
\small
    a_n+b_n+t_n=r_n
\end{equation}
The ambiguity of pair $p_n$ can be calculate as follow:
\begin{equation}
\small
    Ambiguity_{p_n}=\frac{|a_n - b_n|}{r_n}=\frac{|a_n - b_n|}{a_n+b_n+t_n}
\end{equation}
It is straightforward to observe that the ambiguity of $p_n$ lies between $0$ and $1$:\textbf{Ambiguity = 0:} This pair is highly ambiguous, with equal votes for $A$ and $B$, or all participants voting for a tie. \textbf{Ambiguity = 1:} This pair is not ambiguous at all, with all participants voting exclusively for $A$ or $B$.

\subsection{Agreement}
The agreement of observer $o_i$ with all other observers for $p_n$ can be computed with weighted agreement:
\begin{equation}
\small
    W_i(p_n) = \begin{cases}  P(A)_n=a_n/r_n & \text{if } v_{i} = A \\ P(B)_n=b_n/r_n & \text{if } v_i = B \\ P(T)_n=t_n/r_n & \text{if } v_i = T \end{cases}
\end{equation}

where $v_i$ is the vote of observer $o_i$ for video pair $p_n$.

\subsection{Inter-observer Consistency}
For a given observer $o_i$, the basic idea to compute his/her consistency is to calculate the average of the agreement between him/her and other observers across all the pairs he/she voted on. The higher this agreement, the more consistent this observer is with others.
However, computing consistency solely based on agreement is not entirely fair, particularly for videos with high ambiguity. To address this, we weight the agreement based on ambiguity, assigning larger weights to less ambiguous pairs and smaller weights to more ambiguous pairs.
In active sampling, pairs are rated by varying numbers of observers, so the number of ratings per pair must be considered. Pairs with more votes should be weighted higher, while those rated only once should be excluded from consistency calculations.
The inter-observer consistency $C_i$ can be calculated as:
\begin{equation}
\small
    C_{i} = \frac{\sum_{n=1}^{N}(r_n-1)\times\frac{|a_n - b_n|}{r_n}\times W_i(p_n)}{\sum_{n=1}^{N}(r_n-1)}
\end{equation}
The larger $C_i$ is, the more consistent this observer is with others. To further validate the proposed method, we injected synthetic spammers, and detailed results can be found in Sec.~\ref{subsec:screening}.

\section{Resolution Cross-over Quality Loss}
\begin{table*}[t]
\caption{Benchmark RCQL on LSCO datasets}
\label{tab:RCQL}
\vspace{-0.2cm}
\begin{adjustbox}{width=1.8\columnwidth,center}
\begin{tabular}{@{}l|llllllllll@{}}
\toprule
 &
   &
  PSNRy &
  SSIM &
  MS-SSIM &
  VMAF &
  VMAF 4k &
  \begin{tabular}[c]{@{}l@{}}VMAF\\ 4k neg\end{tabular} &
  P1204.3 &
  EQM NR &
  EQM FR \\ \midrule
\multirow{3}{*}{$\Delta$Bitrate (Kbps)$\downarrow$} &
  2160p vs 1080p &
  3175.97 &
  4533.55 &
  3512.11 &
  2930.42 &
  3158.60 &
  3142.59 &
  2472.82 &
  \textbf{1691.42} &
  2042.62 \\
 & 1080p vs 720p & 1028.79         & 5852.18 & 2370.15 & \textbf{845.64} & 847.35 & 977.67 & 1299.44         & 1692.27 & 1886.10 \\
 & 720p vs 540p  & \textbf{308.04} & 2255.98 & 671.32  & 340.51          & 367.33 & 350.80 & 523.37          & 453.48  & 267.61  \\ \midrule
\multirow{3}{*}{$RCQL_{s}$ (JOD$\times$Kbps)$\downarrow$} &
  2160p vs 1080p &
  1861.03 &
  1440.17 &
  1395.24 &
  1635.23 &
  1792.11 &
  1787.83 &
  1234.36 &
  \textbf{487.68} &
  635.71 \\
 & 1080p vs 720p & \textbf{169.67} & 3157.68 & 850.07  & 199.28          & 188.66 & 184.61 & 299.53          & 679.59  & 1189.30 \\
 & 720p vs 540p  & \textbf{32.12}  & 1474.12 & 202.78  & 39.46           & 53.24  & 45.96  & 119.12          & 120.18  & 45.34   \\ \midrule
\multirow{3}{*}{$RCQL_{avg}$ (JOD)$\downarrow$} &
  2160p vs 1080p &
  0.2237 &
  0.1716 &
  0.1746 &
  0.1921 &
  0.2051 &
  0.2047 &
  0.1850 &
  \textbf{0.1091} &
  0.1634 \\
 & 1080p vs 720p & 0.1482          & 0.5510  & 0.2619  & 0.1822          & 0.1711 & 0.1685 & \textbf{0.1423} & 0.2894  & 0.3289  \\
 & 720p vs 540p  & 0.0893          & 0.3750  & 0.1707  & \textbf{0.0770} & 0.0842 & 0.0815 & 0.1781          & 0.1818  & 0.1186  \\ \bottomrule
\end{tabular}
\end{adjustbox}
\vspace{-0.5cm}
\end{table*}
\label{sec:RCQL}
Evaluating VQM performance often involves measuring their correlation (e.g., PLCC, SROCC) with subjective scores (e.g., MOS, JOD) on study datasets. However, higher correlation does not guarantee better resolution cross-over accuracy, as correlation reflects global agreement but overlooks specific errors at the cross-over point.
\begin{figure}[b]
    \centering
    \vspace{-0.5cm}
    \includegraphics[clip,width=0.35\textwidth]{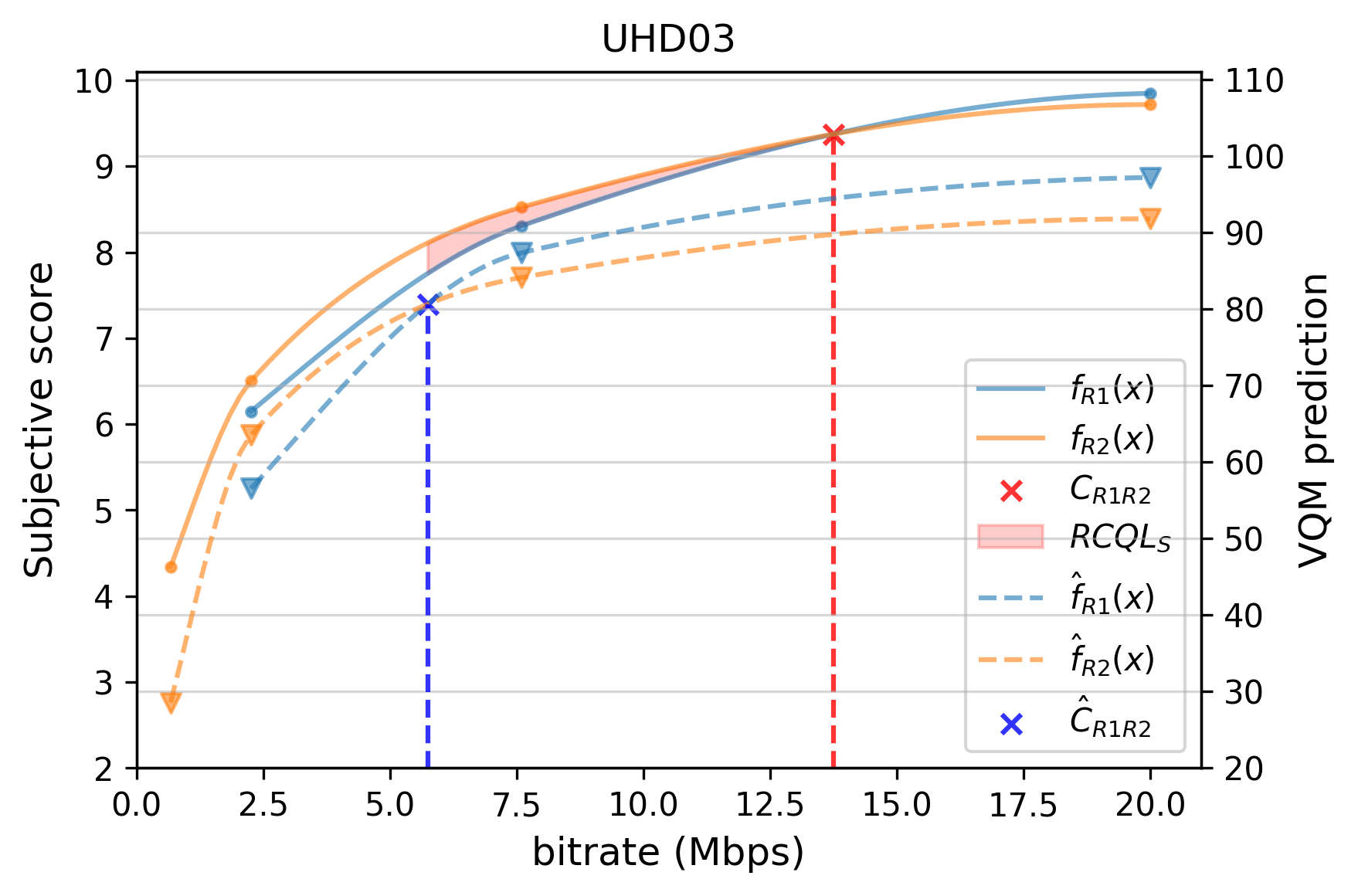}
    \caption{Illustration of RCQL computation for a VQM between two resolutions, $R1$ and $R2$.}
    \label{fig:RCQL_math}
\end{figure}
Fig.~\ref{fig:RCQL_math} illustrates how the RCQL is computed. Suppose there are two resolutions, $R1$ and $R2$, and we aim to determine their resolution cross-over point. The subjective quality is measured, and a function is fitted to describe the corresponding RD curves, denoted as $f_{R1}(x)$ and $f_{R2}(x)$ for the two resolutions, respectively. We assume that both $f_{R1}(x)$ and $f_{R2}(x)$ are convex and monotonically increasing. Similarly, the predicted video quality metric functions are denoted as $\hat{f}_{R1}(x)$ and $\hat{f}_{R2}(x)$. The bitrate at the cross-over point obtained from the subjective study can be computed as:
\begin{equation}
\small
{C_{R1R2}} = \min \{ x\left| {{f_{R1}}(x) = {f_{R2}}(x)} \right|\} 
\end{equation}
Similarly, the bitrate at the cross-over point predicted by the VQM can be computed as:
\begin{equation}
\small
{{\hat C}_{R1R2}} = \min \{ x\left| {{{\hat f}_{R1}}(x) = {{\hat f}_{R2}}(x)} \right|\} 
\end{equation}
The difference between the bitrates of the two cross-over points can be calculated as: 
\begin{equation}
\small
    \Delta Bitrate = \left|C_{R1R2}-\hat{C}_{R1R2} \right|
\end{equation}

It can be observed in Fig.~\ref{fig:RCQL_math} that in the bitrate range between $\hat{C}_{R1R2}$ and $C_{R1R2}$, the subjective study indicates that $R2$ provides better quality than $R1$ and should be selected on the convex hull. However, due to errors in cross-over prediction by the VQM, $R1$ is incorrectly selected instead of $R2$ within this bitrate range. The resulting quality loss for users can be quantified as the integral of the subjective quality difference between $R1$ and $R2$ over this bitrate range. Mathematically, the RCQL is computed as:
\begin{equation}
\small
RCQL = \Bigg| \big| \int_{C_{R1R2}}^{\hat{C}_{R1R2}} f_{R1}(x) \, dx \big| - \big| \int_{C_{R1R2}}^{\hat{C}_{R1R2}} f_{R2}(x) \, dx \big| \Bigg|
\end{equation}
A higher RCQL value indicates lower cross-over accuracy for the given metric.

\begin{figure}[b]
    \centering
    \begin{subfigure}{.24\textwidth}
        \centering
        \includegraphics[width=\textwidth]{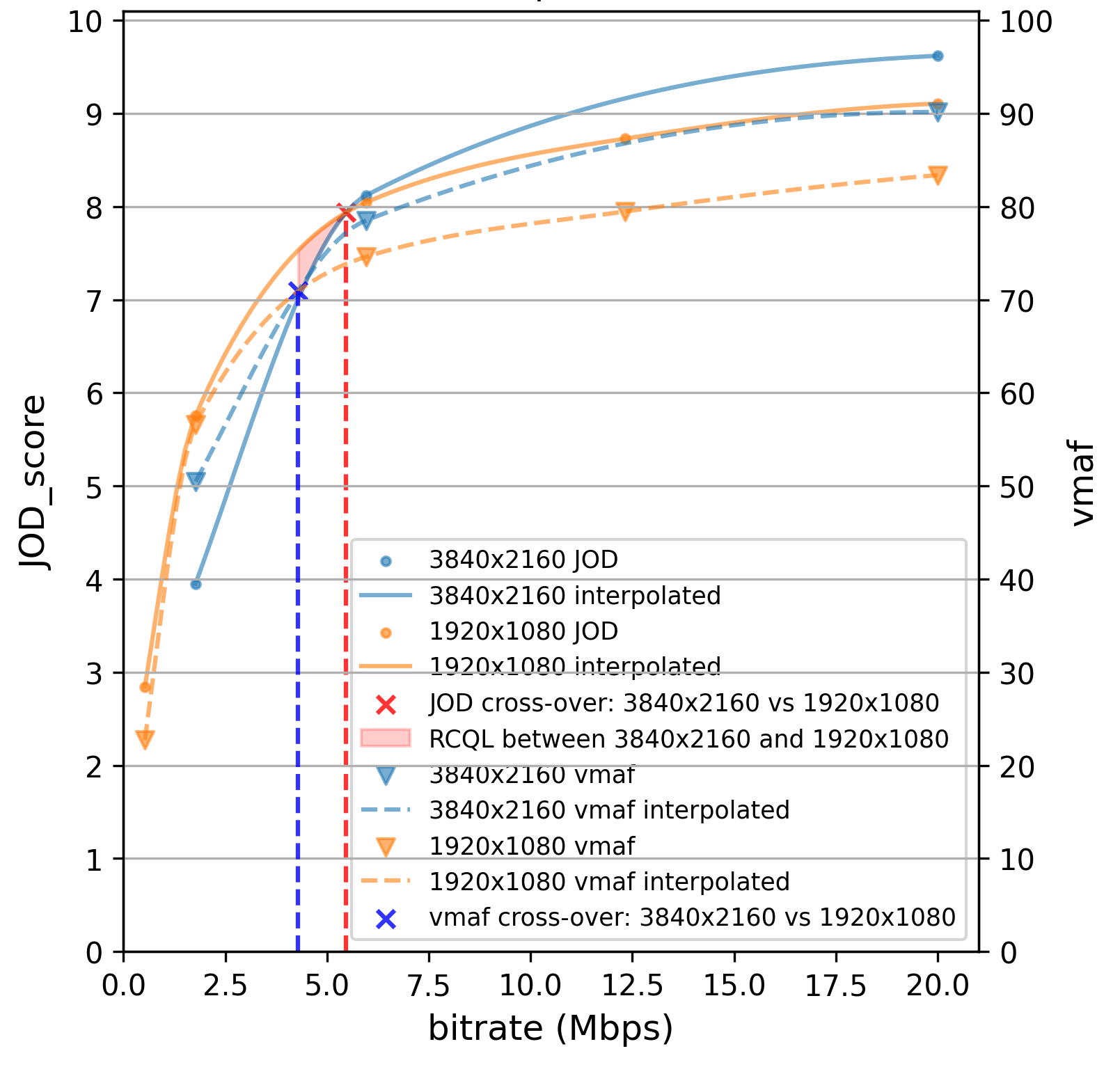}
        \vspace{-0.5cm}
        \caption{Case 1 (EPL5)}
        \label{subfig:RCQL1}
    \end{subfigure}
    \begin{subfigure}{.24\textwidth}
        \centering
        \includegraphics[width=\textwidth]{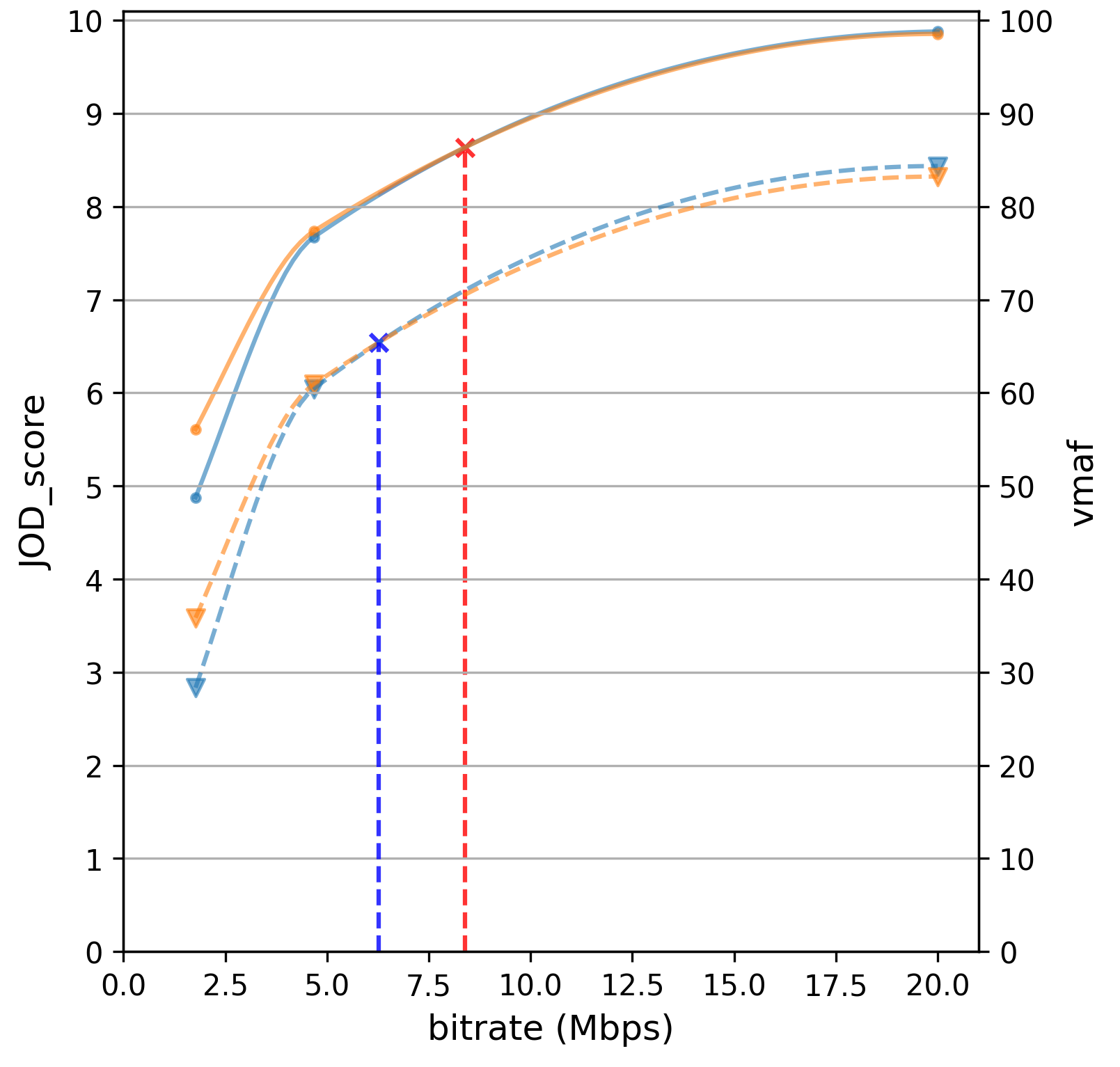}
        \vspace{-0.5cm}
        \caption{Case 2 (UCL1)}
        \label{subfig:RCQL2}
    \end{subfigure}
    \caption{Demonstration of RCQL and $\Delta Bitrate$ between VMAF predictions and subjective scores for the resolution cross-over between 2160p and 1080p.}
    \vspace{-0.5cm}
    \label{fig:RCQL}
\end{figure}

Fig.~\ref{fig:RCQL} demonstrates why $\Delta Bitrate$ alone is not sufficient to measure cross-over accuracy. While the $\Delta Bitrate$ in Case 1 is smaller than in Case 2, the subjective quality difference between $R1$ and $R2$ is much smaller in Case 2 compared to Case 1. As a result, the actual quality loss in Case 2 is lower than in Case 1.

Our proposed RCQL metric effectively measures the quality loss caused by selecting the incorrect resolution. It provides a more accurate and reliable measure to reflect the QoE.
We also computed the average quality loss over the range of the mistake as:
\begin{equation}
\small
    RCQL_{avg}=RCQL_s/\Delta Bitrate
\end{equation}
This reflects the average quality loss (measured in JOD units) over the mistaken bitrate range.
\section{Experimental results}

\subsection{Inter-observer consistency}
\label{subsec:screening}
\begin{figure}[htb]
    \centering  
    \includegraphics[width=0.5\textwidth]{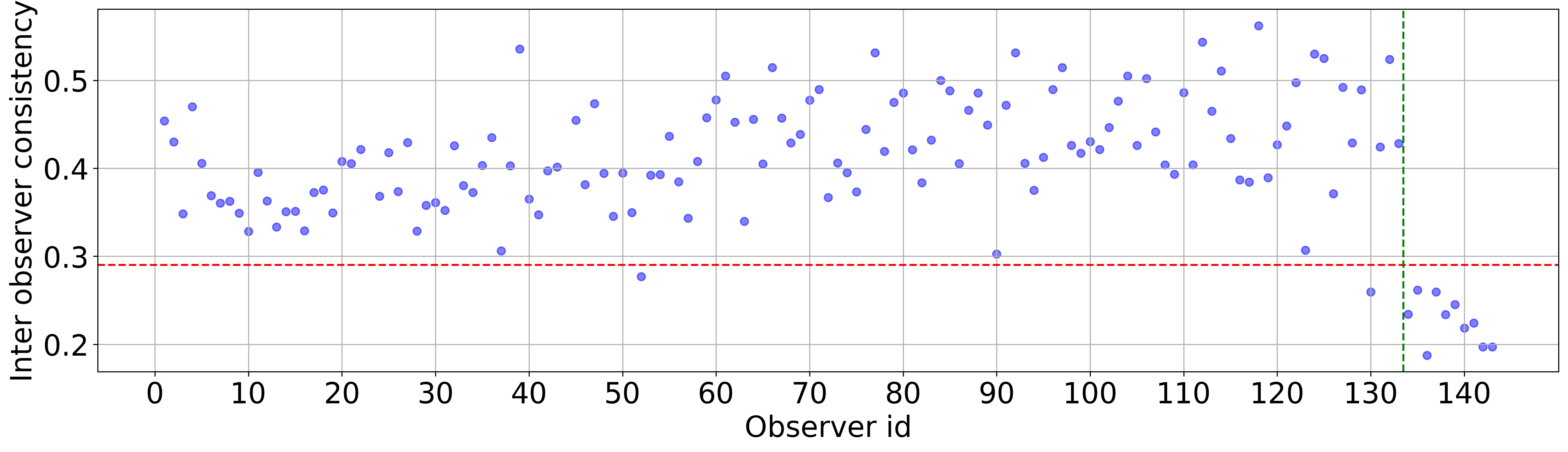}
    \caption{Inter-observer consistency for each participant, including 10 synthetic spammers  (id from 133 to 143)}
    \vspace{-0.2cm}
    \label{fig:spammer}
\end{figure}

We generated 10 synthetic spammers who provided random ratings and computed the inter-observer consistency together with all participants using the method proposed in Sec.~\ref{sec:obs_screen}. The results are presented in Fig.~\ref{fig:spammer}. It can be observed that the synthetic spammers (after the green line) exhibit significantly lower inter-observer consistency. This observation also helps establish a threshold to identify outlier observers. In this study, we chose 0.3 as our threshold and removed 2 outliers before proceeding with further analysis.

\subsection{Benchmark of VQM on our collected LSCO datasets}
We compared various VQMs, including PSNR, SSIM~\cite{wang2004image}, MS-SSIM~\cite{wang2003multiscale}, VMAF~\cite{vmaf}, P1204.3~\cite{rao2020bitstream}, and EQM~\cite{chen2024encoder}. The EQM include a No Reference (NR) model and Full Reference (FR) model.
The correlation results are presented in Table~\ref{tab:correlation}. It can be observed that EQM outperforms other metrics on the entire dataset as well as across different resolutions.
\begin{table}[htb]
\caption{Benchmark of correlation between VQM and our collected LSCO datasets}
\label{tab:correlation}
\begin{adjustbox}{width=1\columnwidth,center}
\begin{tabular}{@{}lllllllll@{}}
\toprule
                       &         & PSNRy  & SSIM   & MS-SSIM & VMAF   & P1204.3 & EQM NR          & EQM FR          \\ \midrule
\multirow{5}{*}{SROCC} & 2160p   & 0.8322 & 0.9091 & 0.8322  & 0.8601 & 0.9371  & \textbf{0.972}  & \textbf{0.972}  \\
                       & 1080p   & 0.7147 & 0.8765 & 0.8206  & 0.8471 & 0.9176  & \textbf{0.9824} & 0.9765          \\
                       & 720p    & 0.7059 & 0.7941 & 0.7118  & 0.8    & 0.8647  & 0.9382          & \textbf{0.9588} \\
                       & 540p    & 0.522  & 0.5714 & 0.4341  & 0.6264 & 0.6978  & \textbf{0.8571} & 0.8462          \\
                       & overall & 0.8149 & 0.8011 & 0.813   & 0.8839 & 0.9321  & \textbf{0.9463} & 0.9457          \\ \midrule
\multirow{5}{*}{PLCC}  & 2160p   & 0.7611 & 0.8371 & 0.7797  & 0.8334 & 0.8765  & 0.9422          & \textbf{0.9504} \\
                       & 1080p   & 0.7381 & 0.846  & 0.7758  & 0.8408 & 0.9378  & 0.9776          & \textbf{0.9801} \\
                       & 720p    & 0.7356 & 0.7357 & 0.7231  & 0.8392 & 0.8949  & \textbf{0.9781} & 0.9778          \\
                       & 540p    & 0.5506 & 0.6173 & 0.5494  & 0.6912 & 0.7827  & \textbf{0.9145} & 0.9074          \\
                       & overall & 0.7831 & 0.7146 & 0.7359  & 0.8558 & 0.9076  & 0.9525          & \textbf{0.954}  \\ \bottomrule
\end{tabular}
\end{adjustbox}
\end{table}
We also evaluated various VQMs on cross-over accuracy using the RCQL metric proposed in Section~\ref{sec:RCQL}. The results in Table~\ref{tab:RCQL} show that different VQMs perform variably across resolutions. For the cross-over between 2160p and 1080p, EQM NF outperforms other metrics. Interestingly, for lower resolution cross-overs, PSNR achieves better performance. 

This observation is interesting because it is commonly known that PSNR has a relatively low correlation with subjective study datasets compared to learning-based VQMs such as VMAF and EQM (as also shown in Table~\ref{tab:correlation}). However, for resolution cross-over accuracy, PSNR is outperforming both VMAF and EQM on low resolution.
This might be because EQM is trained on datasets that focus more on higher resolutions (HD and 4K). Another possible reason is that the resolution cross-over accuracy does not evaluate the cross-content generalization ability of a VQM, which is penalized when using correlation metrics such as SROCC across the entire dataset. This reveals that a higher correlation does not necessarily guarantee better cross-over accuracy.

\section{Conclusion}
In this paper, we demonstrated the limitations of ACR in predicting resolution cross-overs and showed that PC can improve cross-over accuracy. We proposed a mathematical measure of inter-observer consistency to identify and remove spammers during subjective studies. We further introduced the RCQL metric to specifically evaluate cross-over accuracy and benchmarked state-of-the-art VQMs on the LSCO dataset under live streaming scenarios. Experimental results show that higher correlation with subjective scores does not necessarily translate to better cross-over accuracy. Additionally, for resolution cross-overs, EQM performs better at higher resolutions, while PSNR is more effective at lower resolutions. These findings provide valuable insights for selecting appropriate quality metrics when optimizing bitrate ladders in adaptive streaming applications.

\bibliographystyle{IEEEbib}
\bibliography{icme2025references}

\vspace{12pt}

\end{document}